# DNA mechanical deformations and chiral spin selectivity

Solmar Varela,[1, 2] Vladimiro Mujica,[3] and Ernesto Medina[4, 5]

[1] *Yachay Tech, School of Chemical Sciences & Engineering, 100119-Urcuquí, Ecuador*
[2] *Escuela de Física, Facultad de Ciencias, Universidad Central de Venezuela, Caracas, Venezuela*
[3] *Department of Chemistry and Biochemistry, Arizona State University, Tempe, Arizona, 85287, USA*
[4] *Yachay Tech, School of Physical Sciences & Nanotechnology, 100119-Urcuquí, Ecuador*
[5] *Centro de Física, Instituto Venezolano de Investigaciones Científicas (IVIC), Apartado 21827, Caracas 1020 A, Venezuela*
(Dated: October 19, 2017)

The strength of the spin-orbit interaction relevant to transport in a low dimensional structure depends critically on the relative geometrical arrangement of current carrying orbitals. Recent tight-binding orbital models for spin transport in DNA-like molecules, have surmised that the band spin-orbit coupling arises from the particular angular relations between orbitals of neighboring bases on the helical chain. Such arrangement could be probed by inducing deformations in the molecule in a conductive probe AFM type setup, as it was recently reported by Kiran, Cohen and Naaman[1]. Here we report deformation dependent spin selectivity when a double strand DNA model is compressed or stretched. We find that the equilibrium geometry is not optimal with respect to the SO coupling strength and thus spin selectivity can be tuned by deformations. The latter can be increased by stretching the helical structure taking into account its elastic properties through the Poisson ratio. The spin filtering gap is also found to be tunable with uniaxial deformations.

## I. INTRODUCTION

The Chiraly Induced Spin Selectivity (CISS) consists of the strong spin polarization of electrons when they are transmitted through a chiral structure. The CISS effect has been measured in a great variety of chiral molecular structures including single molecules of DNA [2–4], Photosystem I [5], self-assembled monolayers of DNA, chiral oligopeptides[6, 7] and helicenes [8]. As these molecular systems lack strong exchange interactions and magnetic centers, It was first proposed that the spin active ingredients to the photo-electron spin polarization setup[2] was the spin-orbit (SO) coupling[9] in addition to the chiral potential. Such couplings have been considered in recent theoretical models in order to describe transport of electrons through chiral molecules [10–13], where the SO interaction has been added ad-hoc, without specifying in detail, the particular source of the interaction. Furthermore, some models[14, 15] also include time reversal symmetry breaking features by either introducing coupling asymmetries between different conducting channels or adding dephasing by way of Buttiker voltage probe attachments. Both of these ingredients also couple strongly to spin[16].

Recently, in ref. [17] the SO coupling involved in transport is explicitly derived from a tight-binding model of DNA. The coupling is built from the overlap of $p$ orbitals between vicinal bases and its magnitude depends of the DNA helix parameters (radius and pitch) and the relative positions of the orbitals. The magnitude of the SO interaction is derived ultimately from the atomic coupling from a perturbative treatment that connects nearest neighbor sites. Building on these results, we show here that it is possible to tune the magnitude of the SO interaction in a DNA helix model through changes in the radius and pitch due to the longitudinal deformations. Such deformations also modify the gap that protects spin polarized states during the transport process[3]. Such manipulation has been shown in recent experiments with oligopeptides[1] where the capacity to filter spin was changed by compression in a conductive probe AFM set up (see also resistance studies of DNA with stretching[18]).

This work is organized as follows; In section II we derived the effective overlaps between nearest neighbor orbitals of a simple model of DNA. The effective overlaps are a result of the presence of the atomic spin-orbit coupling, using the Slater-Koster formulation[19, 20] and a lowest-order matrix perturbation theory as proposed in[21]. The effective SO coupling obtained is equal to that of reference [10] and depends explicitly on parameters that define the structural configuration of the DNA helix. The analysis of the behavior of the strength of the SO interaction under effects of longitudinal stretching and Poisson ratio effects are shown in Section III. The gap protecting spin transport reported in [10] is also modulated by the structural parameters. It is shown that the conditions for CISS are not optimal for the relaxed structure and can be enhanced by deformations. Furthermore, the detailed deformation dependence found can shed light on the orbitals involved in transport, making the CISS a spectral probe. We close with a summary and conclusions.

## II. ANALYTICAL SLATER-KOSTER MODEL FOR DOUBLE STRANDED DNA

We model double helix DNA as a helical sequence of steps (bases) whose plane is perpendicular to the molecular axis. The bases are planar ring-like structures with in-plane sigma-bonding ( $sp^2$ ) and out of plane orbitals ($p_z$ like) most likely to provide itinerant electrons[22]. Slater-koster overlaps are considered between vicinal bases on each strand. The coupling between strands is considered very weak due to the relatively long distances implied by the hydrogen bonds. The direct overlap between neighbouring $p_z$ orbitals does not couple spin so additional couplings between the bare local orbitals must be considered. The atomic SO interaction couples the electron bearing $p_z$ orbitals to the sigma structure of the planar base, which in turn couples to the neighboring bases. In ref. [10] all lowest order couplings involving SO and external electric fields were considered as a function of the structural parameters of the double helix. The effective coupling were derived by solving the system of coupled equations for the elemental overlaps by a one-step decimation [23] procedure equivalent to the lowest-order perturbation.

Here we derive the effective Hamiltonian of the DNA incorporating spin coupling from the more compact approach of band folding[21, 24]. We consider an intrinsic atomic SO coupling (associated to C or N for the DNA bases) given by

$$H_{\rm SO} = \frac{e\hbar}{4m_0^2 c^2} \mathbf{s} \cdot (\mathbf{p} \times \nabla \mathbf{V}), \qquad (1)$$

where $m_0$ is the effective electron mass, $V$ is the atomic potential, $\mathbf{s}$ is the electron spin, $\hbar$ is Planck's constant, $e$ the electron charge and $c$ is the speed of light. Considering the bare $p$ orbitals on the bases, the possible SO matrix elements between these orbitals are

|  | $\|p_x\rangle$ | $\|p_y\rangle$ | $\|p_z\rangle$ |
|---|---|---|---|
| $\langle p_x\|$ | 0 | $-i\boldsymbol{s}_z \xi_p$ | $i\boldsymbol{s}_y \xi_p$ |
| $\langle p_y\|$ | $i\boldsymbol{s}_z \xi_p$ | 0 | $-i\boldsymbol{s}_x \xi_p$ |
| $\langle p_z\|$ | $-i\boldsymbol{s}_y \xi_p$ | $i\boldsymbol{s}_x \xi_p$ | 0 |

where $\xi_p = \lambda \hbar^2/2$. For carbon atoms $\xi_p \sim 6$ meV. $\mathbf{s}_i = \hbar/2\boldsymbol{\sigma}_i$, with $\boldsymbol{\sigma}_i$ the Pauli matrices in the local orbital system.

The eigenvalue equation for the coupled Hamiltonian is given by

$$\begin{pmatrix} H_\gamma & T \\ T^\dagger & H_\chi \end{pmatrix} \begin{pmatrix} \gamma \\ \chi \end{pmatrix} = E \begin{pmatrix} \gamma \\ \chi \end{pmatrix}. \qquad (2)$$

Here $H_\gamma$ is the sub-space that contains the $p_z$ orbital site energies and the off-diagonal overlaps $E_{zz}$ between the $p_z$ orbitals on $\imath$ and $\jmath$ sites

$$H_\gamma = \begin{pmatrix} \epsilon_{2p}^\pi & E_{zz}^{\imath\jmath} \\ E_{zz}^{\jmath\imath} & \epsilon_{2p}^\pi \end{pmatrix}. \qquad (3)$$

where $\epsilon_{2p}^\pi$ are the bare energies of 2p C levels. The $T$ sub-space contains the intrinsic coupling between the $p_z$ orbitals with orbitals $p_x$ and $p_y$ of the sigma bonded carbon of the base and $E_{xz}^{\imath\jmath}$ and $E_{yz}^{\imath\jmath}$ overlaps between these orbitals on $\imath$ and $\jmath$ sites,

$$T = \begin{pmatrix} 0 & -i\boldsymbol{s}_y\xi_p & i\boldsymbol{s}_x\xi_p & 0 & E_{zx}^{\imath\jmath} & E_{zy}^{\imath\jmath} \\ 0 & E_{zx}^{\jmath\imath} & E_{zy}^{\jmath\imath} & 0 & -i\boldsymbol{s}_y\xi_p & i\boldsymbol{s}_x\xi_p \end{pmatrix}. \qquad (4)$$

Finally, the $H_\chi$ sub-space contains, the energies $\epsilon_s$, $\epsilon_{2p}^\sigma$ and $\epsilon_{2p}^\sigma$ of the $s$, $p_{x,y}$ and $p_z$ orbitals of the nearest neighbor base.

$$H_\chi = \mathrm{diag}[\epsilon_s, \epsilon_{2p}^\sigma, \epsilon_{2p}^\sigma, \epsilon_s, \epsilon_{2p}^\sigma, \epsilon_{2p}^\sigma]. \qquad (5)$$

The wave function subspaces $\gamma = (\psi_{\imath p_z}, \psi_{\jmath p_z})$ and $\chi = (\psi_s, \psi_{p_x}, \psi_{p_y})$ are coupled by $T$.

The matrix elements $E_{\mu\mu'}^{\sigma,\pi}$ that represent the overlaps of bare orbitals are expressed as a linear combination of Slater-Koster parameters $V_{\mu\mu'}^{\sigma,\pi}$ which are related to the different types of molecular bonding ($\sigma$, $\pi$) between the atomic wavefunctions of the $\mu$ and $\mu'$ orbitals[19, 25]. These elements obey the commutation rule $V_{ll'} = (-1)^{l+l'} V_{l'l}$ where $l$ is the orbital angular momentum quantum number ($l=1$ for $p$ orbitals). Harrison et al [25] proposes an empirical expression for orbital overlaps as a function of the interatomic distance $\mathbf{R}_{\jmath\imath}$, given that

$$V_{\mu\mu'}^{\sigma,\pi} = \kappa_{\mu\mu'(\sigma,\pi)} \frac{\hbar^2}{m_e R_{\jmath\imath}^2}, \qquad (6)$$

where $\kappa_{\mu\mu'(\sigma,\pi)}$ depends on the particular atom and

$$|\mathbf{R}_{\jmath\imath}|^2 = \frac{16\pi^2 a^2 \sin^2(\Delta\phi/2) + (b\Delta\phi)^2}{4\pi^2}. \qquad (7)$$

If separation of the atoms is large, the dependence on the distance is exponential.

As the bases rotate along the helix we must consider that the orbitals do not have the same absolute orientation at each site[20]. Using $\hat{\mathbf{X}}, \hat{\mathbf{Y}}, \hat{\mathbf{Z}}$ as the basis fixed in space, we define the unit vectors $\hat{\mathbf{n}}(\mu_\jmath)$ in the direction the orbital $\mu_\jmath$ living on the helix as

$$\hat{\mathbf{n}}(p_x, \imath) = \cos\phi_\imath \hat{\mathbf{X}} + \sin\phi_\imath \hat{\mathbf{Y}}, \qquad (8)$$

$$\hat{\mathbf{n}}(p_y, \imath) = \cos\phi_\imath \hat{\mathbf{X}} + \sin\phi_\imath \hat{\mathbf{Y}}, \qquad (9)$$

$$\hat{\mathbf{n}}(p_z, \imath) = \hat{\mathbf{Z}}, \qquad (10)$$

with $\phi_\imath = (\imath - 1)\Delta\phi$, where $\imath = 1...N$ and $N$ is the total number of sites on helix. If the orbitals $\mu_\imath$ are located in $\mathbf{R}_\imath$ and the orbitals $\mu'$ are in $\mathbf{R}_\jmath$, overlap between the orbitals at $\imath$ and $\jmath$ sites are given by

$$E_{\mu\mu'}^{\imath,\jmath} = (\hat{\mathbf{n}}(\mu_\imath), \hat{\mathbf{n}}(\mu'_\jmath)) V_{\mu\mu'} \qquad (11)$$

$$+ \frac{(\mathbf{R}_{\jmath\imath}, \hat{\mathbf{n}}(\mu_\imath))(\mathbf{R}_{\jmath\imath}, \hat{\mathbf{n}}(\mu'_\jmath))}{(\mathbf{R}_{\jmath\imath}, \mathbf{R}_{\jmath\imath})} \left(V_{\mu\mu'}^\sigma - V_{\mu\mu'}^\pi\right), (12)$$

where $\mathbf{R}_{\jmath\imath} = \mathbf{R}_\jmath - \mathbf{R}_\imath$ the vector connecting the $\imath$ and $\jmath$ sites. By Eq. (12) we obtain that matrix elements are

$$E_{zz}^{\imath\jmath} = V_{pp}^\pi + \frac{b^2(\Delta\phi)^2 (V_{pp}^\sigma - V_{pp}^\pi)}{4\pi^2 |\mathbf{R}_{\jmath\imath}|^2} = E_{zz}^{\jmath\imath}, \qquad (13)$$

$$E_{xz}^{\imath\jmath} = \frac{2ab\nu \sin^2\left(\frac{\Delta\phi}{2}\right)\Delta\phi\, (V_{pp}^\sigma - V_{pp}^\pi)}{2\pi |\mathbf{R}_{\jmath\imath}|^2} = -E_{zx}^{\imath\jmath}, \qquad (14)$$

where $\nu_{ij} = sgn(j-i)$ and

$$E_{yz}^{ij} = \frac{ab\Delta\phi \sin\Delta\phi \left(V_{pp}^{\sigma} - V_{pp}^{\pi}\right)}{2\pi|\mathbf{R}_{ji}|^2} = E_{zy}^{ij}. \quad (15)$$

The dependencies of the overlaps on the geometric properties of the helix i.e. pitch $b$, rotation angle between bases $\Delta\phi$, and helix radius $a$ will parameterize the molecular deformations and determine the coupling strengths.

Decimating the subspace $\chi$ and retaining up to linear terms in $E$ and lowest order in the coupling $T$ (2), we obtain that effective Hamiltonian by the relation[21, 24]

$$H = S^{-1/2}\left[H_{\gamma} - TH_{\chi}^{-1}T^{\dagger}\right]S^{-1/2}, \quad (16)$$

where $S = 1 + TH_{\chi}^{-1}T^{\dagger}$.

The product $TH_{\chi}^{-1}T^{\dagger}$ is expanded as

$$TH_{\chi}^{-1}T^{\dagger} = \begin{pmatrix} 0 & -i s_y \xi_p & i s_x \xi_p & 0 & E_{zx}^{ij} & E_{zy}^{ij} \\ 0 & E_{zx}^{ji} & E_{zy}^{ji} & 0 & -i s_y \xi_p & i s_x \xi_p \end{pmatrix} \begin{pmatrix} \frac{1}{(\epsilon_{2p}^{\pi}-\epsilon_s)} & 0 & 0 & 0 & 0 & 0 \\ 0 & \frac{1}{(\epsilon_{2p}^{\pi}-\epsilon_{2p}^{\sigma})} & 0 & 0 & 0 & 0 \\ 0 & 0 & \frac{1}{(\epsilon_{2p}^{\pi}-\epsilon_{2p}^{\sigma})} & 0 & 0 & 0 \\ 0 & 0 & 0 & \frac{1}{(\epsilon_{2p}^{\pi}-\epsilon_s)} & 0 & 0 \\ 0 & 0 & 0 & 0 & \frac{1}{(\epsilon_{2p}^{\pi}-\epsilon_{2p}^{\sigma})} & 0 \\ 0 & 0 & 0 & 0 & 0 & \frac{1}{(\epsilon_{2p}^{\pi}-\epsilon_{2p}^{\sigma})} \end{pmatrix} \begin{pmatrix} 0 & 0 \\ i s_y \xi_p & E_{xz}^{ij} \\ -i s_x \xi_p & E_{yz}^{ij} \\ 0 & 0 \\ E_{xz}^{ji} & i s_y \xi_p \\ E_{yz}^{ji} & -i s_x \xi_p \end{pmatrix}. \quad (17)$$

Replacing Eq. (17) into Eq. (16) and approximating $S \sim 1$, the effective hamiltonian is

$$H \approx H_{\gamma}' - \begin{pmatrix} \frac{2\xi_p^2 + (E_{zx}^{ij})^2 + (E_{zy}^{ij})^2}{(\epsilon_{2p}^{\pi}-\epsilon_{2p}^{\sigma})} & \frac{2i\xi_p s_y E_{zx}^{ij}}{(\epsilon_{2p}^{\pi}-\epsilon_{2p}^{\sigma})} \\ \frac{-2i\xi_p s_y E_{zx}^{ij}}{(\epsilon_{2p}^{\pi}-\epsilon_{2p}^{\sigma})} & \frac{2\xi_p^2 + (E_{zx}^{ij})^2 + (E_{zy}^{ij})^2}{(\epsilon_{2p}^{\pi}-\epsilon_{2p}^{\sigma})} \end{pmatrix}, \quad (18)$$

were $H_{\gamma}'$ is the sub-space with $\epsilon_{2p}^{\pi} = 0$, where we have used the symmetry relations

$$\begin{aligned} E_{xz}^{ij} &= -E_{zx}^{ij} = -E_{xz}^{ji}, \\ E_{yz}^{ij} &= E_{zy}^{ij} = E_{yz}^{ji}. \end{aligned} \quad (19)$$

Terms on the diagonal of (18) give both corrections to the $p_z$ orbital energy and the effective coupling between $p_z$ orbitals. Using expresions (13)-(7), the Hamiltonian for the full Brillouin zone can be written as

$$H = t \sum_{\langle ij \rangle} c_i^{\dagger} c_j + i\lambda_{\rm SO} \sum_{\langle ij \rangle} c_i^{\dagger} \nu_{ij} \mathbf{s}_y c_j, \quad (20)$$

where $\nu_{ij} = sgn(j-i)$ and

$$t = E_{zz} = V_{pp}^{\pi} + \frac{b^2 \Delta\phi^2 \left(V_{pp}^{\sigma} - V_{pp}^{\pi}\right)}{8\pi^2 a^2(1-\cos\Delta\phi) + b^2 \Delta\phi^2} \quad (21)$$

is the kinetic term, and

$$\lambda_{\rm SO} = \frac{8\pi\xi_p ab\Delta\phi \sin^2\left(\frac{\Delta\phi}{2}\right)\left(V_{pp}^{\sigma} - V_{pp}^{\pi}\right)}{\left(\epsilon_{2p}^{\pi} - \epsilon_{2p}^{\sigma}\right)\left(16\pi^2 a^2 \sin^2\left(\frac{\Delta\phi}{2}\right) + b^2 \Delta\phi^2\right)}, \quad (22)$$

is the effective intrinsic SO coupling, recovering the expressions derived in ref.[17]. Here the two helices are uncoupled so we will have two identical copies of the same Hamiltonian translating into two channels for transport. In order to estimate the contribution of $V_{pp}^{\sigma} - V_{pp}^{\pi}$ we will use Eqs.6 and 7.

## III. STRETCHING EFFECTS ON SO COUPLING

In this section we will derive the effects, on the SO strength, of feasible experimental deformations. As shown in ref.[1], *conductive probe AFM* can be used to tip-load an oligopeptide and modulate its spin filtering capacity. Different types of loadings could reveal further interesting features for the orbital overlaps involved in spin-filtering and most interestingly the source of transport spin-orbit coupling. In our simple model, to consider stretching or compression in DNA we assume that the orbitals on the bases do not change their orientation and $\Delta\phi$ remains invariant during the deformation process. On the other hand we consider that the dsDNA has a Poisson ratio as reported in experiments [26], which is our simplistic account for the mechanical behavior of the two chains bonded together.

For a helix DNA with $\mathcal{N}$ turns and $N$ bases in total, the length of the chain can be written as

$$L = \frac{(N-1)\Delta\phi}{2\pi}b = \mathcal{N}b. \quad (23)$$

Using the Poisson's ratio for DNA equal to $\nu$ and considering a longitudinal deformation $\varepsilon = \Delta L/L_o$, SO interaction changes as

$$\lambda_{\text{SO}}(\Delta L) = \frac{32\pi^3 \xi_p (\kappa_{pp}^\sigma - \kappa_{pp}^\pi) \left(a_o - \nu a_o \frac{\Delta L}{L_o}\right) \left(b_o + \frac{\Delta L}{\mathcal{N}}\right) \Delta\phi \sin^2\left(\frac{\Delta\phi}{2}\right)}{(\epsilon_{2p}^\pi - \epsilon_{2p}^\sigma) \left(16\pi^2 (a_o - \nu a_o \frac{\Delta L}{L_o})^2 \sin^2\left(\frac{\Delta\phi}{2}\right) + (b_o + \frac{\Delta L}{\mathcal{N}})^2 \Delta\phi^2\right)^2}, \quad (24)$$

where $a_o$ is the radius and $L_o$ is the length of the DNA molecule without stretching and $\Delta L$ is the change of length during deformation. If $\Delta L < 0$ ($\Delta L > 0$) the molecule is compressed (stretched) in the longitudinal direction and radius increases (decreases) See Fig.1. The ratio $(\kappa_{pp}^\sigma - \kappa_{pp}^\pi)/(\epsilon_{2p}^\pi - \epsilon_{2p}^\sigma)$ is a constant only relating to the nature of the bonded orbitals.

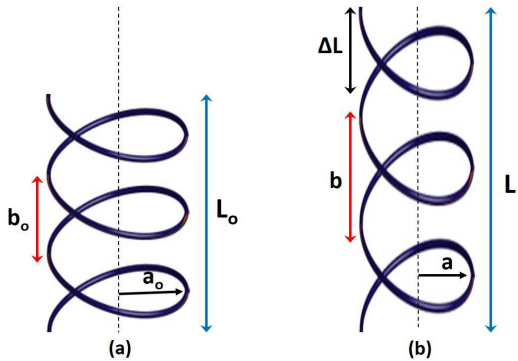

FIG. 1. (a) Schematic illustration of the deformation of the DNA helix of radius $a_o$ and pitch $b_o$. (b) After $\Delta L$ deformation, the radius and pitch change as $a = a_o - \nu a_o \Delta L/L_o$ and $b = b_o + \Delta L/\mathcal{N}$, respectively.

Figure 2 displays the SO magnitude versus the longitudinal deformation $\varepsilon$. $\varepsilon$ varies in interval $-1$ to $1/\nu$. We emphasize some particular features of the overlaps: When $\varepsilon = -1$, the helix is completely compressed, the pitch $b$ being zero and the resulting structure is a ring with $a = a_0(1 + \nu)$. In this case, the SO magnitude is zero because the $E_{xz}^{ij}$ overlaps disappear killing the SO interaction. On the other hand, when $\varepsilon$ approaches $1/\nu$, the pitch $b$ is large compared to the helix radius $a$, the SO interaction weakens because overlaps depend on the inverse of the separation between the bases. At a particular value for $\varepsilon$ the SO coupling is a maximum as a function of $\Delta L$ where $b$ and $a$ are coupled by the Poisson ratio.

The Poisson ratio for dsDNA is reported to be $\nu = 0.5$ in experiments[26]. For this value $\varepsilon = 0.57$ yields and maximum for the SO magnitude and its value is approximately 52% greater than the value without deformation. Note that compressing the double helix would yield a decrease of the SO coupling. These behaviors could be tested in detailed experiments, to verify the origin of the SO coupling and the orbital overlaps involved in transport.

Nevertheless, the Fig.2 shows a broader range of stretching than can actually be achieved without breaking the molecule. According to measurements performed on DNA, molecules between 5 and 25 base pairs can only be stretched 1.4 to 1.6 Å. To show a more realistic range for stretching, we depict the SO coupling changes in the inset of Fig.2.

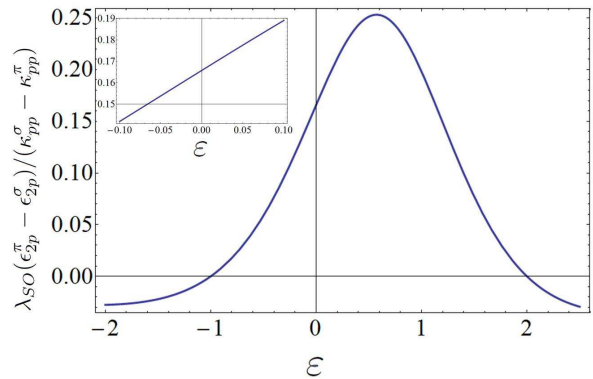

FIG. 2. $\lambda_{\text{SO}}$ normalized by $(\kappa_{pp}^\sigma - \kappa_{pp}^\pi)/(\epsilon_{2p}^\pi - \epsilon_{2p}^\sigma)$ versus the deformation $\varepsilon$ of a DNA helix. The plot shows that stretching can enhance the SO coupling, while compressing weakens the coupling. The inset shows a more realistic range for $\varepsilon$ according to reference [18].

## IV. EFFECTS OF DNA DEFORMATION ON THE SPIN FILTERING GAP

From the Hamiltonian described in Eq.20 one can derive the corresponding Bloch Hamiltonian. Our model of DNA assumes that transport electrons are available on the $2p - \pi$ orbitals of the bases. These orbitals have one unpaired electron, so we choose to describe the Bloch physics of the half filled model. The corresponding Hamiltonian in the vicinity of this point is

$$H = -2\nu q_y R t - 2\nu \lambda_{\text{SO}} \boldsymbol{s}_y, \quad (25)$$

where $q_y$ is the wave-vector measuring the separation from half filling in reciprocal space in the rotating frame with respect to the fixed $\hat{\mathbf{X}}, \hat{\mathbf{Y}}, \hat{\mathbf{Z}}$ reference frame (see Eq.10), $R$ is given by Eq.7 and $\nu$ is a quantum number denoting the sense of rotation of electrons on the helix. Writing the operator $\boldsymbol{s}_y$ in the rotating frame one arrives at the Hamiltonian [17]

$$H_{helix} = \nu \begin{pmatrix} iT\partial_\varphi & 2i\lambda_{\text{SO}} e^{-i\varphi} \\ -2i\lambda_{\text{SO}} e^{i\varphi} & iT\partial_\varphi \end{pmatrix}, \quad (26)$$

written in spin space in the rotating frame, where $T = 2Rt$. The eigenvalues associated with the energy are given by

$$E_{n,s}^{\nu,\zeta} = \begin{cases} \frac{|T|n}{2M\mathcal{N}}, & \lambda_{\text{SO}} = 0 \\ \frac{|T|n}{2M\mathcal{N}} - s\nu\frac{\sqrt{T^2 + (4\lambda_{\text{SO}})^2}}{2}, & \lambda_{\text{SO}} \neq 0 \end{cases}, \quad (27)$$

with $\mathcal{N}$ the number of turns in the helix, $M$ the number of bases per turn, $n$ that gives the subbands corresponding to the discrete modes due to longitudinal confinement within a helix [27] and $s\nu$ is the helicity of the electron associate to the spin component $s$. Note the singular behavior due to different symmetries of models with and without the SO coupling in the limit $\lambda_{\text{SO}} \to 0$. The two cases have to be derived separately. This model predicts

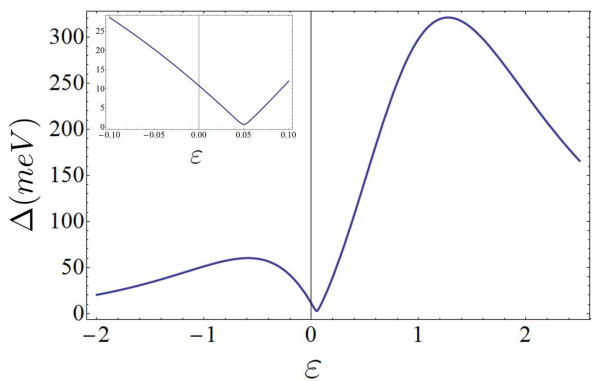

FIG. 3. Spin filtering gap versus deformation of a helix of dsDNA. The figure shows the general trend for a wide range of values of $\varepsilon$ while the inset emphasizes the physical range of deformations according to ref.[18]. We have taken the values of $\kappa_{pp\sigma} = 3.24$ $\kappa_{pp\pi} = -0.81$ from ref.[25].

a gap $\Delta$ that separates the different helicities for the same direction of transport given by

$$\Delta = |T|\left(\sqrt{1 + \left(\frac{4\lambda_{\text{SO}}}{T}\right)^2} - \frac{1}{2\mathcal{N}}\right), \quad (28)$$

for $\lambda_{\text{SO}} \neq 0$. If $\lambda_{\text{SO}} = 0$ then $\Delta = 0$ and the states of different helicity are degenerate as expected. As shown in ref.[17] this gap protects spin states as in topological insulators and is a critical ingredient for spin filtering once time-reversal is broken by e.g. an external bias. In Figure 3 we show the gap $\Delta$ given by Eq.28 as a function of molecular stretching. We note that the dependencies reflected in Fig.3 result from both changes in $T$ (through $t$ in Eq.21), the kinetic term, and the SO coupling. The gap gets a much more pronounced increase when stretching as opposed to compressing (with a slight offset). The latter offset becomes important when looking at a physical range parameters according to ref. [18], as shown in the inset of Fig.3. The gap has a non-monotone dependence on stretching while it can be enhanced by compressing the molecule.

This gap is also sensitive to the number of turns in the molecule, as shown in Fig.4.

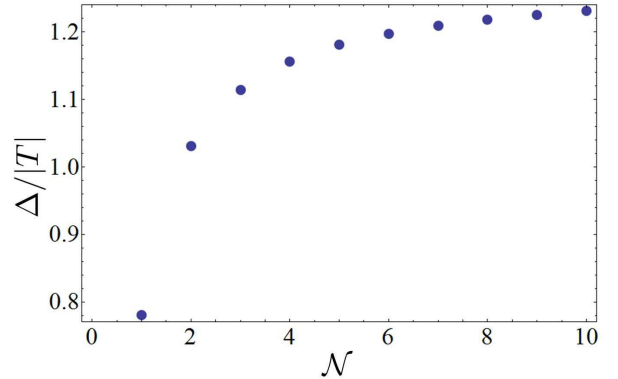

FIG. 4. Spin filtering gap in units of the kinetic energy $|T|$ versus the length of the helix. The energy gap separating the spin up states from spin down states, for a given helicity, increases with the length of the helix for fixed $a$ and $b$.

## V. SUMMARY AND CONCLUSIONS

We have developed an analytical model for the SO coupling in dsDNA assuming transport electrons are provided by $p_z$ orbitals from $2p$ levels that are projected perpendicular to the base molecules. The source of the SO coupling is atomic, associated to Carbon and Nitrogen atoms in the base. We derived the molecular Hamiltonian by using the Slater-Koster tight-binding approach on a rotating frame. The spin-orbit coupling is derived explicitly in terms of the atomic SO interaction (one orbital per base) and the orbital overlaps between nearest neighbor bases along each helix of double stranded DNA. Coupling between strands is only mechanical and no transport between strands is considered. Having the SO coupling as a function of the geometrical parameters of the dsDNA we compute the changes in the interaction by longitudinal deformations of the molecule. The relation between pitch and radius as stretching occurs is controlled by the experimentally reported Poisson ratio of dsDNA.

We find that, according to our simple model, stretching dsDNA can increase the SO coupling by at least 10% while compressing uniaxially (no bending effects) reduces the interaction. We also derive the dependence of the spin filtering gap discussed in ref.[17] and the its dependence on stretching and length of the dsDNA. The gap is affected both by changes in the kinetic term and the SO coupling. We find that while stretching reduces the gap, compressing can increase its value by a factor of two. Changes in the gap are more revealing of the filtering capacity of the molecule than the values of the SO coupling alone. Experiments using conductive probe AFM[1, 18] could directly verify our results and help settle the ques-tion of the source of the SO coupling and the conduction electrons in dsDNA. Finally, it is important to note an observation of Brout et al[18] on the distribution of deformation of a helical molecule based on the De Gennes elastic model[28]: induced deformations will be non-uniform and concentrated at the ends of the molecule. A feature to contemplate in future detailed modelling of stretching effects on spin transport.

## ACKNOWLEDGMENTS

We thank Henry Pinto for illuminating discussions. This work was supported by an internal grant of Yachay Tech University.